\documentclass[12pt]{article}
\usepackage{graphicx}
\usepackage{amsmath}
\usepackage{graphics}
\usepackage{epstopdf, color}

\usepackage{hyperref}
\begin{document}
{\bf \Large
\begin{center}
{Massive fermions interacting via a harmonic oscillator in the presence of a minimal length uncertainty relation}
\end{center}}
\vspace{4mm}
\begin{center}
{{\large{\bf B. J. Falaye $^{a,b,\dag,}$}\footnote{\scriptsize E-mail:~ fbjames11@physicist.net;~ babatunde.falaye@fulafia.edu.ng\\ \dag{Corresponding} author}\large{\bf ,} {\large{\bf Shi-Hai Dong $^{c,}$}}\footnote{\scriptsize E-mail:~ dongsh2@yahoo.com}}}\large{\bf,} {\large{\bf K. J. Oyewumi $^{d,}$\footnote{\scriptsize E-mail:~ kjoyewumi66@unilorin.edu.ng}}\large{\bf ,} {\large{\bf K. F. Ilaiwi $^{e,}$}}\footnote{\scriptsize E-mail:~ khaled.ilawi@najah.edu} \large{\bf and} {\large{\bf S. M. Ikhdair $^{e,f,}$}}\footnote{\scriptsize E-mail:~ sameer.ikhdair@najah.edu;~ sikhdair@gmail.com.}}
\end{center}

{\footnotesize
\begin{center}
{\it $^{\textbf{a}}$ESFM, Instituto Polit\'ecnico Nacional, UPALM, M\'{e}xico D. F. 07738, M\'{e}xico.}\\
{\it $^\textbf{b}$Applied Theoretical Physics Division, Department of Physics, Federal University Lafia,  P. M. B. 146, Lafia, Nigeria.}\\
{\it $^\textbf{c}$CIDETEC, Instituto Polit\'{e}cnico Nacional, UPALM, M\'{e}xico D. F. 07700, M\'{e}xico.}\\
{\it $^\textbf{d}$Theoretical Physics Section, Department of Physics, University of Ilorin,  P. M. B. 1515, Ilorin, Nigeria.}\\
{\it $^\textbf{e}$Department of Physics, Faculty of Science, An-Najah National University, P. O. Box 7, Nablus, West Bank, Palestine.}\\
{\it $^\textbf{f}$Department of Electrical Engineering, Near East University, Nicosia, Northern Cyprus, Mersin 10, Turkey.}
\end{center}}

\textbf{
\begin{center}
Int. J. Mod. Phys. E {\bf24} (2015) 1550087
\end{center}
}

\noindent
\begin{abstract}
\noindent
We derive the relativistic energy spectrum for the modified Dirac equation by adding a harmonic oscillator potential where the coordinates and momenta are assumed to obey the commutation relation $\left[\hat{x},\hat{p}\right]=i\hbar\left(1+\eta p^2\right)$. In the nonrelativistic limit, our results are in agreement with the ones obtained {previously}.
Furthermore, the extension to the construction of creation and annihilation operators for the harmonic oscillators with minimal length uncertainty relation is presented. Finally, we show that the commutation relation of the $su(1, 1)\sim so(2,1)$ algebra is satisfied by the {operators $\hat{\mathcal{L}_{\pm}}$ and $\hat{\mathcal{L}_{z}}$}.
\end{abstract}



\section{Introduction}
In recent years, the study of the harmonic oscillator problem in connection with the minimal length formalism {has} attracted the interest of theoretical physicists [1-16]. This minimal length may be described in a quantum system theoretically as a nonzero minimal uncertainty in position measurements. A major feature of this concept is that the physics below such a scale becomes inaccessible and then it defines a natural cut-off which prevents the usual UV divergences \cite{N2}. This interesting subject has been studied within the frameworks of various branches of physics including string theory \cite{N12}, quantum cosmology \cite{N13}, black hole physics \cite{N14}, general relativity \cite{N15} and special relativity \cite{N16,N17}.

In 2002, Chang et al. \cite{NW5}  illustrated how certain features of string theory may manifest themselves in simple quantum mechanical systems through the modification of the canonical commutation relation. Recently, Pedram \cite{N11} presented the energy eigenvalues of a quantum bouncer in the framework of the Generalized (Gravitational) Uncertainty Principle (GUP) via quantum mechanical and semiclassical schemes. He also used two equivalent nonperturbative representations of a deformed commutation relation in the {form $[\hat x,\hat p]=i\hbar(1+\eta p^2)$}, where $\eta$ is the GUP parameter. This new representation is formally self-adjoint and preserves the ordinary nature of the position operator. He showed that both representations lead to the same modified semiclassical energy spectrum and agree well with the quantum mechanical description. Very recently, the Nikiforov-Uvarov technique has been applied to solve the Dirac equation in minimal length quantum mechanics \cite{N18}.

{Using the formula method for bound state problems \cite{BJ4} in the present work, we determine {the solutions} for fermionic massive {spin $1/2$} particles, interacting with a harmonic oscillator, where the coordinates and momenta are assumed to obey the commutation relation $\left[\hat{x},\hat{p}\right]=i\hbar\left(1+\eta p^2\right)$. Furthermore, we obtain the nonrelativistic limit of our solution and then proceed to the construction of the creation and annihilation operators for harmonic oscillator with minimal length uncertainty relation}.

\section{Formulation of the problem}
The Heisenberg algebra we need to study, can be obtained when the commutation relation between the position and momentum is modified from the canonical one to \cite{N11}
\begin{equation}
\left[\hat{x},\hat{p}\right]=i\hbar\left(1+\eta p^2\right),
\label{EE1}
\end{equation}
where $\eta$ is an extremely small positive deformation parameter. According to {this algebra,} which implements the minimal length, we have the deformed uncertainty relation, which appears in perturbative string theory as
\begin{equation}
{\Delta x\geq\frac{\hbar}{2}\left(\frac{1}{\Delta p}+\eta\Delta p\right)=\frac{\hbar}{2}\left[\left(\frac{1}{\sqrt{\Delta p}}-\sqrt{\eta\Delta p}\right)^2+2\sqrt{\eta}\right]},
\label{EE2}
\end{equation}
and this implies the existence of a minimal length
\begin{equation}
\left(\Delta x\right)_{min}=\hbar\sqrt{\eta}.
\label{EE3}
\end{equation}

We denote the position and momentum operators obeying equation (\ref{EE1}) in momentum space by \cite{NW5}
\begin{equation}
\hat{x}=i\hbar\left[\left(1+\eta p^2\right)\frac{\partial}{\partial p}+\gamma p\right],\ \ \ \ \ \ \hat{p}=p.
\label{EE4}
\end{equation}
Here $\gamma$ is an arbitrary constant, which does not appear in the commutation relation (1), and does not affect the relativistic energy spectrum we derive, but only the weight function in the scalar product in the momentum space \cite{NW5, BJ2}
\begin{equation}
\left\langle \Xi'|\Xi\right\rangle=\int\frac{dp}{\left[f(p)\right]^{1-\alpha}}\Xi'^*(p)\Xi(p),
\label{EE5}
\end{equation}
with $f(p)=1+\eta p^2$ and {$\alpha=\gamma/\eta$}. Now, consider the Dirac equation of a particle with mass $m$ moving under the influence of multi-scalar $S(x)$ and $V(x)$ potentials as
\begin{equation}
{\left\{\vec{\zeta}\cdot\vec{p}+\beta\left[m+S(\hat{x})\right]\right\}\Psi_n(p)=\left[E_R-V(\hat{x})\right]\Psi_n(p)},
\label{EE6}
\end{equation}
where $E_R$ is the relativistic energy of the system, and {{$\vec{\zeta}$}} and $\beta$ are $4\times 4$ Dirac matrices defined by
\begin{equation}
\vec{\zeta}=\left(\begin{matrix}0&\vec{\sigma}\\ \vec{\sigma}&0\end{matrix}\right),\ \ \ \beta=\left(\begin{matrix}I&0\\0&-I\end{matrix}\right).
\label{EE7}
\end{equation}
Taking the spinor eigenfunctions as
\begin{equation}
\Psi_n(p)=\left(\begin{matrix}\phi_n(p)\\ \Omega_n(p)\end{matrix}\right),
\label{EE8}
\end{equation}
and substituting them into equation (\ref{EE6}), we find the following two-coupled equations
\begin{subequations}
\begin{eqnarray}
\left(\vec{\sigma}\cdot\vec{p}\right)\phi_n(p)-\left(m+S(\hat{x})-V(\hat{x})\right)\Omega_n(p)&=&E_R\Omega_n(p),
\label{EE9a}\\
\left(\vec{\sigma}\cdot\vec{p}\right)\Omega_n(p)+\left(m+S(\hat{x})+V(\hat{x})\right)\phi_n(p)&=&E_R\phi_n(p).
\label{EE9b}
\end{eqnarray}
\end{subequations}
Further solving equations (\ref{EE9a}) and (\ref{EE9b}) simultaneously and making use of the formula
\begin{equation}
{\left(\vec{\sigma}\cdot\vec{A}\right)\left(\vec{\sigma}\cdot\vec{B}\right)=\left(\vec{A}\cdot\vec{B}\right)+i\vec{\sigma}\cdot\vec{A}\times\vec{B}},
\label{EE10}
\end{equation}
we can eliminate $\Omega_n(p)$ and then obtain
\begin{equation}
p^2\phi_n(p)+\left(S^2(\hat{x})-V^2(\hat{x})+2mS(\hat{x})+2E_RV(\hat{x})\right)\phi_n(p)=\left(E_R^2-m^2\right)\phi_n(p).
\label{EE11}
\end{equation}
For the case of an equal multi-scalar potential, we obtain the following wave equation\footnote{It is worth mentioning that only the choice $S(\hat{x}) = +V(\hat{x})$ produces a nontrivial nonrelativistic limit with a potential function $2V(\hat{x})$, and not $V(\hat{x})$. Accordingly, it would be natural to scale the  potential terms in equation (\ref{EE9a}) and equation (\ref{EE9b}) so that in the nonrelativistic limit the  interaction potential becomes $V(x)$, not $2V(x)$ \cite{BJ3,BJ3b}. Thus, these modifications have been made in equation (\ref{EE12}).}
\begin{equation}
p^2\phi_n(p)+V(\hat{x})\left(E_R+m\right)\phi_n(p)=\left(E_R^2-m^2\right)\phi_n(p).
\label{EE12}
\end{equation}
Now inserting the harmonic oscillator potential, {$V(\hat{x})=m\omega^2{\hat{x}}^2/2$} into equation (\ref{EE12}) we get
\begin{equation}
p^2\phi_n(p)+\frac{m\omega^2{\hat{x}}^2}{2}\left(E_R+m\right)\phi_n(p)=\left(E_R^2-m^2\right)\phi_n(p).
\label{EE13}
\end{equation}
Furthermore, substituting operator  (\ref{EE4}) into equation (\ref{EE13}) and using
\begin{equation}\label{EE14}
\hat{x}^2=(-\hbar^2)\left[\gamma  \left(p^2 (\eta +\gamma )+1\right)\phi (p)+2 p (\eta +\gamma ) \left(\eta  p^2+1\right) \phi '(p)+\left(\eta  p^2+1\right)^2 \phi ''(p)\right]
\end{equation}
as well as defining the following two parameters
\begin{equation}
\tilde{A}=\frac{2}{\hbar^2m\omega^2\left(E_R+m\right)}-\gamma(\gamma+\eta)\ \ \ \ \mbox{and}\ \ \ \tilde{B}=-\left(\frac{2\left(E_R^2-m^2\right)}{\hbar^2m\omega^2(E_R+m)}+\gamma\right),
\label{EE15}
\end{equation}
we finally obtain
\begin{equation}
\frac{d^2\phi_n(p)}{dp^2}+\frac{2(\gamma+\eta)p}{(1+\eta p^2)}\frac{d\phi_n(p)}{dp}-\frac{\tilde{B}+p^2\tilde{A}}{(1+\eta p^2)^2}\phi_n(p)=0.
\label{EE16}
\end{equation}
We employ the formula method (FM) to solve equation (\ref{EE16}). The FM has been introduced in Ref. \cite{BJ4} and has been used in some physical systems to obtain the whole spectra in a very simple way. Here we proceed with its application to relativistic equations in minimal length quantum mechanics. We give a brief review of this method and all formulas required for our calculation in the next section.

\section{Method of analysis}
One of the calculational tools utilized in solving the Schr\"{o}dinger-like equation including the centrifugal barrier and/or the spin-orbit coupling term is the so called FM. For a given potential the idea is to convert the Schr\"{o}dinger-like wave equation into the standard form given by equation (1) of ref. \cite{BJ4}, i.e.
\begin{equation}
\Psi''(s)+\frac{(k_1-k_2s)}{s(1-k_3s)}\Psi'(s)+\frac{(As^2+Bs+C)}{s^2(1-k_3s)^2}\Psi(s)=0,
\label{EE17}
\end{equation}
via an appropriate coordinate transformation of the form $s=s(r)$. {The prime represents the first derivative with respect to the argument $s$, and $A,B,C$ are parameters}. Then, the energy eigenvalues and the corresponding wave functions can be obtained  if the problem is exactly solvable. We use the following formulas
\begin{subequations}
\begin{eqnarray}
&&\left[\frac{k_4^2-k_5^2-\left[\frac{1-2n}{2}-\frac{1}{2k_3}\left(k_2-\sqrt{(k_3-k_2)^2-4A}\right)\right]^2}{2\left[\frac{1-2n}{2}-\frac{1}{2k_3}\left(k_2-\sqrt{(k_3-k_2)^2-4A}\right)\right]}\right]^2-k_5^2=0, k_{3} \neq 0 \ \ \ \mbox{and}
\label{EE18A}\\[2mm]
&&\Psi_n(s)=N_ns^{k_4}(1-k_3s)^{k_5}\ _2F_1\left(-n,\ n+2(k_4+k_5)+\frac{k_2}{k_3}-1;\ 2k_4+k_1,\ k_3s\right),
\label{EE18B}
\end{eqnarray}
\end{subequations}
respectively, where
\begin{equation}
k_4=\frac{1-k_1+\sqrt{(1-k_1)^2-4C}}{2},\ \
k_5=\frac{1}{2}+\frac{k_1}{2}-\frac{k_2}{2k_3}+\sqrt{\left[\frac{1}{2}+\frac{k_1}{2}-\frac{k_2}{2k_3}\right]^2-\left[\frac{A}{k_3^2}+\frac{B}{k_3}+C\right]},
\label{EE19}
\end{equation}
and $N_{n}$ is the normalization constant.

\section{Exact solution to the problem}
In order to solve equation (\ref{EE16}) by the FM, we introduce a new transformation of the {form $\varsigma=\arctan\left(p\sqrt{\eta}\right)/\sqrt{\eta}$ with the interval  $-\pi/(2\sqrt{\eta})<\varsigma<\pi/(2\sqrt{\eta})$}, which maintains the finiteness of the transformed wave functions on the boundaries. Thus, we find
\begin{equation}
\frac{d^2\phi_n(\varsigma)}{d\varsigma^2}+\frac{2\gamma}{\sqrt{\eta}}\tan(\varsigma\sqrt{\eta})\frac{d\phi_n(\varsigma)}{d\varsigma}-\left(\tilde{B}+\frac{\tilde{A}\tan^2(\varsigma\sqrt{\eta})}{\eta}\right)\phi_n(\varsigma)=0.
\label{EE20}
\end{equation}
Furthermore, we perform a substitution $\varsigma\rightarrow\varrho=\sin(\varsigma\sqrt{\eta})$  which maps the {region $(-\pi/2\sqrt{\eta}<\varsigma<\pi/2\sqrt{\eta})$ to $-1<\varrho<1$}. In that case, equation (\ref{EE20}) can be transformed into
\begin{equation}
\frac{d^2\phi_n(\varrho)}{d\varrho^2}+\left(\frac{2\gamma}{\eta}-1\right)\left(\frac{\varrho}{1-\varrho^2}\right)\frac{d\phi_n(\varrho)}{d\varrho}
+\frac{\varrho^2(\tilde{B}\eta-\tilde{A})-\tilde{B}\eta}{\eta^2(1-\varrho^2)^2}\phi_n(\varrho)=0.
\label{EE21}
\end{equation}
Now, let us transform this equation to the form of equation (\ref{EE17}). To this end, we introduce another transformation of the {form $s=(1-\varrho)/2$} to obtain
\begin{equation}
\frac{d^2\phi_n(s)}{ds^2}-\left(\frac{\gamma}{\eta}-\frac{1}{2}\right)\left(\frac{1-2s}{s(1-s)}\right)\frac{d\phi_n(s)}{ds}
+\frac{4s^2(\tilde{B}\eta-\tilde{A})-4s(\tilde{B}\eta-\tilde{A})-\tilde{A}}{4\eta^2s^2(1-s)^2}\phi_n(s)=0.
\label{EE22}
\end{equation}
Since equation (\ref{EE22}) is now suitable for an FM solution, we compare it with equation (\ref{EE17}) and thus, we can determine the parameters $k_{i}$ ($i=1,2,3$) {where $k_1=-\left(\gamma/\eta-1/2\right)$, $k_2 = -2\left(\gamma/\eta-1/2\right)$ and $k_3=1$. Also, $A = (\tilde{B}\eta-\tilde{A})/\eta^2=-B$ and $C = -\tilde{A}/4\eta^2$. Furthermore, by means of equation (\ref{EE19}), we can derive $k_4$ and $k_5$ as follows:}
\begin{eqnarray}
{k_4=k_5=\frac{1}{4}+\frac{\gamma}{2\eta}+\frac{1}{2}\sqrt{\frac{1}{4}+\frac{2}{m\omega^2\eta^2\hbar^2\left(E_R+m\right)}}=\tilde{v}}.
\label{EE23}
\end{eqnarray}
Using the equation (\ref{EE18A}) which can be rewritten as
\begin{equation}
k_4+k_5=\frac{1-2n}{2}-\frac{1}{2k_3}\left(k_2-\sqrt{(k_3-k_2)^2-4A}\right),
\label{EE24}
\end{equation}
leads to the following relativistic energy spectrum with minimal length uncertainty relation
\begin{equation}
\frac{2\left(E_R-m\right)}{\hbar^2\eta m\omega^2}-(2n+1)\sqrt{\frac{1}{4}+\frac{2}{\hbar^2\eta^2m\omega^2\left(E_R+m\right)}}-\frac{1}{4}=\left(\frac{1}{2}+n\right)^2,
\label{EE25}
\end{equation}
or
\begin{equation}
E_R=m+\frac{\hbar\omega m}{2}\left[(2n+1)\sqrt{\frac{\hbar^2\eta^2\omega^2}{4}+\frac{2}{m\left(E_R+m\right)}}+\hbar\eta\omega\left(n^2+n+\frac{1}{2}\right)\right].
\label{EE26}
\end{equation}
Now we calculate the wave functions by employing Eq. (\ref{EE18B}) and find
\begin{equation}
g_n(s)=N_n\left[s(1-s)\right]^{\tilde{v}}\ _2F_1(-n,n+4\tilde{v}-\frac{2\gamma}{\eta}, 2\tilde{v}+\frac{1}{2}-\frac{\gamma}{\eta},  s),
\label{EE27}
\end{equation}
or equivalently
\begin{equation}
g_n(\varrho)=N_n\left(\frac{1-\varrho^2}{4}\right)^{\tilde{v}}\ _2F_1\left(-n,n+4\tilde{v}-\frac{2\gamma}{\eta}, 2\tilde{v}+\frac{1}{2}-\frac{\gamma}{\eta},  \frac{1-\varrho}{2}\right).
\label{EE28}
\end{equation}
Thus the momentum-space wave function takes the form
\begin{eqnarray}
\phi_n(\varrho)=N_n\left(\frac{1-\varrho^2}{4}\right)^{\tilde{v}}{\bf C}_n^{2\tilde{v}-\frac{\gamma}{\eta}}(\varrho)
\label{EE29}
\end{eqnarray}
with {$\varrho=\sin\left(\arctan(p\sqrt{\eta})\right)=p\sqrt{\eta}/\sqrt{1+{\eta}p^2}$}. We see that the momentum-space wave function can also be expressed in terms of Gegenbauer polynomials ${\bf C}_n^\lambda(t)$.

Now, let us take the nonrelativistic limit of our solution. De Souza Dutra et al. \cite{BJ11} found that there is a possibility of obtaining approximate nonrelativistic (NR) solution from relativistic ones. Very recently, Sun \cite{BJ12} proposed a meaningful approach for deriving the bound state solutions of NR Schr\"{o}dinger equation (SE) from the bound state of relativistic equations. The essence of this approach is that, in the NR limit, the SE may be derived from the relativistic one when the energies of the two potentials $S(r)$ and $V(r)$ are small compared to the rest energy $mc^2$, then the NR energy approximated as $E_{n}\rightarrow E-mc^2$ and the NR wave function is $\psi^{NR} (r)\rightarrow\psi(r)$. That is, NR energies, {$E_{n}$} can be determined by taking NR limit values of the relativistic eigenenergies $E_{R}$. Therefore, we can write
\begin{equation}
m+E_R\rightarrow {2\mu}\ \ \ \ \ \ \ \ \  \mbox{and}\ \ \ \ \ \ \ \ \ m-E_R\rightarrow-E_{n},
\label{EE30}
\end{equation}
where $\mu$ is the reduced mass in a NR case \cite{BJ3b}. Consequently, the NR energy spectrum becomes
\begin{equation}
{E_{n}=\hbar\omega\left[\left(\frac{1}{2}+n+n^2\right)\frac{\hbar \mu\eta\omega}{2}+\left(n+\frac{1}{2}\right)\sqrt{\frac{\hbar^2\eta^2\mu^2\omega^2}{4}+1}\right]}
\label{EE31}
\end{equation}
and the momentum-space wave function
\begin{equation}
\phi^{NR}_n(\varrho)=N_n\left(\frac{1-\varrho^2}{4}\right)^{{v}}{C}_n^{\lambda}(\varrho),~~~\lambda= 2{v}-\frac{\gamma}{\eta},~~~ \mbox{with}\ \ \ \ v=\frac{1}{4}+\frac{\gamma}{2\eta}+\frac{1}{2}\sqrt{\frac{1}{4}+\frac{1}{\mu^2\omega^2\eta^2\hbar^2}},
\label{EE32}
\end{equation}
which essentially coincide with the ones obtained previously via the functional analysis approach \cite{NW5}. Before we end this work, we shall identify the creation and annihilation operators for the wavefunctions via the factorization method.
\begin{figure}[!t]
\begin{center}
\includegraphics[height=100mm,width=150mm]{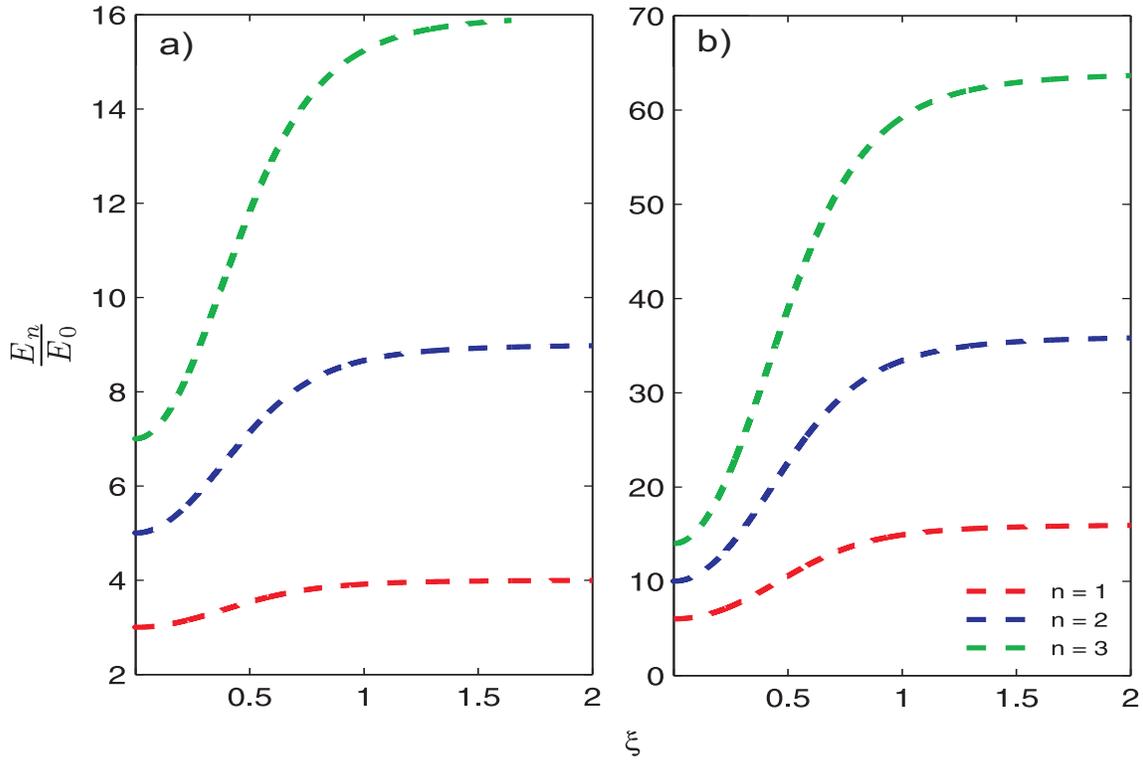}
\end{center}
\caption{{\protect\footnotesize (color online) a) A plot of the energy eigenvalues of the harmonic oscillator in units of the regular ground-state energy as a function of the minimal length in units of the Bohr radius $\xi = \hbar\sqrt{\eta}/a_0$. For principal quantum numbers $n = 1, 2, 3$, ($a_0=1/km$, $\hbar=\mu=1$, $\omega=\pi$). b) Same as a) but with $\omega=2\pi$.}}
\end{figure}

\section{The construction of the ladder operators}
We study the problem of finding the creation and annihilation operators for the position-space wave function which we can obtain by equation (\ref{EE32}) via the FM. As shown in previous works [30-38], the ladder operators can be constructed directly from the wave function by considering the recursion relations of the {special functions} and then constructing an appropriate Lie algebra in terms of these ladder operators without introducing any auxiliary variable. Therefore, we search for differential operators $\mathcal{\hat{L}_{\pm}}$ satisfying the property
\begin{equation}
\mathcal{\hat{L}_{\pm}}\phi_n(p)=l_{\pm}\phi_{n\pm1}(p).
\label{EE33}
\end{equation}
Specifically, these operators are in the form
\begin{equation}
\mathcal{\hat{L}_{\pm}}=A_{\pm}(p)\frac{d}{dp}+ B_{\pm}(p),
\label{EE34}
\end{equation}
which depends only on the physical variable $p$. The calculation in this section is impossible without calculating the normalization constant $N_n$. For this purpose, we utilize equations (\ref{EE32}) and (\ref{EE5}) and get
\begin{equation}
\int_{-\infty}^{\infty}\frac{1}{(1+\eta p^2)^{1-\frac{\gamma}{\eta}}}\left|\phi_n(p)\right|^2dp=N_n^24^{-2v}\int_{-1}^1(1-\varrho^2)^{\lambda-\frac{1}{2}}\left[{\bf C}_n^{\lambda}(\varrho)\right]^2d\varrho=1.
\label{EE35}
\end{equation}
Using the following standard {integral \cite{A38}}
\begin{equation}
\int_{-1}^1(1-x^2)^{t-\frac{1}{2}}\left[C_n^t(x)\right]^2dx=\frac{\pi 2^{1-2t}\Gamma(2t+n)}{n!(n+t)[\Gamma(t)]^2},
\label{EE36}
\end{equation}
we find the normalization constant
\begin{equation}
N_n=\sqrt{\frac{n!\left(n+\lambda\right)\Gamma\left(\lambda\right)^2}{2^{\left(1-2\lambda\right)}\pi\Gamma\left(2\lambda+n\right)}}.
\label{EE37}
\end{equation}
Now, the action of the differential operator ${d}/{d\varrho}$ on wave functions (\ref{EE32}) yields
\begin{equation}
\frac{d}{d\varrho}\phi_n(\varrho)=-\frac{2\varrho v}{1-\varrho^2}\phi_n(\varrho)+N_n4^{-v}(1-\varrho^2)^{v}\frac{d}{d\varrho}{C}_n^{\lambda}(\varrho).
\label{EE38}
\end{equation}
We employ equation (\ref{EE38}) to construct the ladder operators via the recurrence relations of the Gegenbauer polynomials so that we can deduce a relation between $\phi_n(\varrho)$ and $\phi_{n\pm1}(\varrho)$. Differentiating the Gegenbauer polynomial function with respect to $\varrho$ and then utilizing the recursion relation \cite{A38}, we have
\begin{eqnarray}
\frac{d}{d\varrho}C_n^\lambda(\varrho)&=&\frac{n+2\lambda-1}{1-\varrho^2}C_{n-1}^\lambda(\varrho)-\frac{n\varrho}{1-\varrho^2}C_n^\lambda(\varrho)\nonumber\\
&=&\frac{(n+2\lambda)\varrho}{1-\varrho^2}C_{n}^\lambda(\varrho)-\frac{n+1}{1-\varrho^2}C_{n+1}^\lambda(\varrho).
\label{EE39}
\end{eqnarray}
On substituting this expression into equation (\ref{EE38}), we find the equations
\begin{subequations}
\begin{eqnarray}
\frac{d}{d\varrho}\phi_n(\varrho)&=&-\left(\frac{2v\varrho+n\varrho}{1-\varrho^2}\right)\phi_n(\varrho)+\left(\frac{2\lambda+n-1}{1-\varrho^2}\right)\frac{N_n}{N_{n-1}}\phi_{n-1}(\varrho),
\label{EE40}\\
\frac{d}{d\varrho}\phi_n(\varrho)&=&\left(2\lambda-2v+n\right)\frac{\varrho}{1-\varrho^2}\phi_n(\varrho)-\left(\frac{n+1}{1-\varrho^2}\right)\frac{N_n}{N_{n+1}}\phi_{n+1}(\varrho),
\end{eqnarray}
\end{subequations}
and therefore the annihilation operator $\hat{\mathcal{L_{-}}}$ and the creation operator $\hat{\mathcal{L_{+}}}$ are
\begin{subequations}
\begin{eqnarray}
\hat{\mathcal{L_{-}}}=\left[(1-\varrho^2)\frac{d}{d\varrho}+(2v+n)\varrho\right]\sqrt{\frac{\lambda+n-1}{n+\lambda}}\\ \hat{\mathcal{L_{+}}}=\left[-\left(1-\varrho^2\right)\frac{d}{d\varrho}+\left(2\lambda-2v+n\right)\right]\sqrt{\frac{\lambda+n+1}{n+\lambda}},
\label{EE41a}
\end{eqnarray}
with the following properties
\begin{equation}\label{}
\hat{\mathcal{L_{\pm}}}\phi_{n}(\varrho)=l_{\pm}\phi_{n\pm 1}(\varrho),\label{EE41b}
\end{equation}where $l_{\pm}$ are given by
\begin{equation}
l_{-}=\sqrt{n\left(2\lambda+n-1\right)}\ \ \ \mbox{and}\ \ \ \ l_{+}=\sqrt{(n+1)\left(2\lambda+n\right)}.
\label{EE41c}
\end{equation}
\end{subequations}
Let us now calculate the algebra associated with the operators $\hat{\mathcal{L_{\pm}}}=\hat{\mathcal{L}}_{x}\pm i \hat{\mathcal{L}}_{y}$ via equation and (\ref{EE41a})-(\ref{EE41c}). Thus, we determine the commutator
\begin{eqnarray}
\left[\hat{\mathcal{L_{-}}}, \hat{\mathcal{L_{+}}}\right]\phi_n(\varrho)&=&\sqrt{(n+1)\left(2\lambda+n\right)}\left[\hat{\mathcal{L_{-}}}\phi_{n+1}\right]-\sqrt{n\left(2\lambda+n-1\right)}\left[\hat{\mathcal{L_{+}}}\phi_{n-1}\right]\nonumber\\
&=&2\left(\lambda+n\right)\phi_n(\varrho)=2l_0\phi_n(\varrho),
\label{EE42}
\end{eqnarray}
where the eigenvalue $l_0$ has been introduced and the corresponding operator $\hat{\mathcal{L}_{0}}$ is defined as
\begin{equation}
\hat{\mathcal{L}_{0}}=\lambda+\hat{n},
\label{EE43}
\end{equation}
where the number operator $\hat{n}$ \cite{BJ19} has the property $\hat{n}\phi^{NR}_n(\varrho)=n\phi^{NR}_n(\varrho)$. The operators $\hat{\mathcal{L_{\pm}}}$  and $\hat{\mathcal{L}_{0}}$ thus satisfy the commutation relation
\begin{equation}
\left[\hat{\mathcal{L_{-}}},\hat{\mathcal{L_{+}}}\right]=2\hat{\mathcal{L}_{0}}\ ,\ \ \ \left[\hat{\mathcal{L}_{0}},\hat{\mathcal{L_{-}}}\right]=-\hat{\mathcal{L_{-}}}\ ,\ \ \ \left[\hat{\mathcal{L}_{0}},\hat{\mathcal{L_{+}}}\right]=\hat{\mathcal{L_{+}}},
\label{EE44}
\end{equation}
which correspond to the $su(1, 1)$ algebra. The Casimir operator can be written as
\begin{equation}
\hat{\mathcal{C}}\phi_n(\varrho)=\hat{\mathcal{L}_{0}}\left(\hat{\mathcal{L}_{0}}-1\right)\phi_n(\varrho)-\hat{\mathcal{L_{+}}}\hat{\mathcal{L_{-}}}\phi_n(\varrho)=\lambda(\lambda-1)\phi_n(\varrho),
\label{EE46}
\end{equation}
which satisfies the following commutation relation:
\begin{equation}
\left[\hat{\mathcal{C}}, \hat{\mathcal{L}_{\pm}}\right]=\left[\hat{\mathcal{C}}, \hat{\mathcal{L}_{z}}\right]=0.
\label{EE48}
\end{equation}

\section{Conclusion}
The introduction of a minimal observable length was to eliminate divergences that appear in quantum field {theory}. This assumption has led to a generalized uncertainty principle. The consequence of this result is a modification of position and momentum operators according to Eq. (\ref{EE1}). We have studied this in relativistic quantum mechanics by solving the momentum space representation of the modified Dirac equation oscillator via the FM. In the nonrelativistic limit, we have obtained the {results} for a Schr\"odinger system. These results are in agreement with the ones obtained previously in the literature \cite{NW5}.

In Figure 1, we plot the energy eigenvalues of the harmonic oscillator in units of ground state energy versus minimal length in units of Bohr radius $\xi=\hbar\sqrt{\eta}/a_0$. In the usual case when $\xi=0$ or $\eta=0$ the ratio is $E_n/E_0=3, 5,7$ for $n=1,2,3$, respectively. This can be seen from Figure 1 and Eq. (\ref{EE26}) when one sets  $\eta=0$. On the other hand, for any $n$, when  $\xi>1$ the ratio {$E_n/E_0$} remains approximately constant (independent of $\xi$). However, when $\xi<1$, for instance, the ratio $E_n/E_0$ rise sharply for $n=3$ from 7 to 16 times. Therefore, increasing the quantum number $n$ will lead to an increase in that ratio.

Furthermore, the ladder operators for the harmonic oscillator with minimal length uncertainty relation have also been obtained. Finally, we have shown that the commutation relation of the dynamic $su(1, 1)$ algebra is satisfied by the operators $\hat{\mathcal{L}_{\pm}}$, $\hat{\mathcal{L}_{z}}$.

\section*{Acknowledgments}
We thank the kind referees for the positive enlightening comments and suggestions, which have greatly helped us in making improvements to this paper. In addition, BJF acknowledges eJDS (ICTP). This work is partially supported by 20150964-SIP-IPN, Mexico.

\end{document}